\documentclass[aps,twocolumn,amsmath,amssymb,amsfonts,floatfix,showkeys,showpacs,superscriptaddress]{revtex4-1}
\usepackage{dcolumn}  
\usepackage{multirow}
\usepackage[T1]{fontenc}
\usepackage[latin1]{inputenc}
\usepackage{verbatim}
\usepackage{bm} 
\usepackage{hyperref}
\usepackage{natbib}
\usepackage[english]{babel}
\usepackage{babelbib}
\usepackage{graphicx}
\usepackage{epstopdf}
\usepackage{ifpdf} 
\usepackage{epsfig}
\usepackage{color}
\usepackage[usenames,dvipsnames]{xcolor}
\usepackage{mathrsfs}

\begin{document}

\title{Rectification of polymer translocation through nanopores by chiral and nonchiral active particles}

\author{Zahra Fazli}
\thanks{z.fazli@ipm.ir (corresponding author)}
 \affiliation{School of Physics, Institute for Research in Fundamental Sciences (IPM), Tehran 19395-5531, Iran}
 \affiliation{School of Nano Science, Institute for Research in Fundamental Sciences (IPM), Tehran 19395-5531, Iran}
\author{Ali Naji}
\thanks{a.naji@ipm.ir}
 \affiliation{School of Nano Science, Institute for Research in Fundamental Sciences (IPM), Tehran 19395-5531, Iran}
 
\begin{abstract}
We study unbiased translocation of a flexible polymer chain through a membrane pore under the influence of active noise and steric exclusion using Langevin dynamics simulations. The active noise is incorporated by introducing nonchiral and chiral active particles on one or both sides of the membrane. 
Translocation of the polymer into either side of the pore is assisted by an effective pulling due to particle activity and is hindered by an effective pushing due to steric repulsions between the polymer and active particles. 
As a result of the competition between these effective forces, we find a transition between two rectified ({\em cis}-to-{\em trans} and {\em trans}-to-{\em cis}) states. 
This transition is identified by a sharp increase of translocation time. It varies depending on the system parameters such as particle activity, area fraction and chirality whose effects are explored in this work.

\end{abstract}

\maketitle

\section{Introduction}

Micro/nanoscale active particles appear in a wide range of living and artificial examples, ranging from microorganisms such as bacteria and algae to nano/microrobots and synthetic Janus particles \cite{Lauga,Vicsek,Zhang}. They often exhibit self-propelled translational motion in liquid media. This motion which is caused by specific internal mechanisms that make use of the ambient free energy. Such mechanisms may involve ciliary or flagellate organelles as in the case of microorganisms and phoretic forces due to autocatalytic surface reactions in the case of Janus  particles. Active particles can display fascinating collective  properties and self-assembled structures that cannot be understood within the framework of equilibrium physics \cite{Bechinger,Marchetti,Julicher,Elgeti_2,Needleman}. As a particular example, self-assembled colloidal chains with directed motion have been realized using, e.g.,  electrohydrodynamic convection rolls  in nematic liquid crystals \cite{Sasaki}, imbalanced electrostatic interactions in a system of metal-dielectric Janus colloids \cite{Yan} and hydrodynamic interactions due to phoretic flows produced around catalyst-coated colloids linked as a chain \cite{Biswas}. 

Theoretical and computational models of active chains and filaments have been explored in a growing number of recent works; see, e.g., Refs. \cite{Vliegenthart,Loi,Osmanovic,Mousavi,Denk,Lowen,Abaurrea} and references therein. Biological filaments such as F-actin and microtubules in the presence of motility assays (e.g., molecular motors) and also active liquid crystals (active nematics) \cite{Harada,Kumar,Whitfield,Srivastava,Thampi} can be modeled as semiflexible polar active chains being  tangentially propelled \cite{Isele_Holder,Isele_Holder_2}. A computationally suitable model of active polymers is that of harmonically linked active Brownian beads that move in two spatial dimensions (2D)  \cite{Chelakkot,Kaiser}. Various aspects of such active polymers have been studied, including chain (hydro)dynamics and diffusivity in viscous and viscoelastic media \cite{Ghosh,Laskar,Jayaraman,Jiang,Liverpool,Vandebroek}, shape deformations \cite{Eisenstecken,Eisenstecken_2,Kaiser_2}, relaxation under imposed shear, shear-induced alignment and shear thinning \cite{Martin_2,Winkler_2}.

Another emerging scenario involves passive chains or filaments suspended in a bath of active Brownian particles. It is shown that conformational statistics and elastic properties of passive polymers can significantly be altered by the bath activity \cite{Samanta,Kaiser_2,Harder}. A number of peculiar properties such as atypical swelling and looping have been reported in the case of a flexible filament and it is shown that the bath activity can lead to occurrence of a modulational instability \cite{Shin,Nikola}. 

Polymer translocation through nanopores in a dividing membrane has been a major area of interest in soft matter physics (see, e.g., Refs. \cite{Matysiak,Polson,Sakaue,Luo,Chuang,Kantor,Sarabadani,Sung,Muthukumar,Palyulin,Sakaue_2} and references therein). This has partly been due to the significance of polymer translocation to important biological processes and technological applications, such as viral DNA injection \cite{Berndsen,Inamdar}, translocation of single- and double-stranded DNA through nanopores (including $\alpha$-hemolysin and solid-state nanopores) \cite{Wanunu,Maglia}, and DNA sequencing \cite{Branton,Marie}. 

In the context of polymer translocation in active media, effective pulling and pushing forces due to the accumulation of active particles around the polymer and pore are shown to be responsible for nonmonotonic dependence of translocation time on particles activity and their volume fraction \cite{Pu}.
It is shown in Ref. \cite{Khalilian} that self-propelled rods facilitate polymer translocation by inducing a net force, and a scaling relation based on iso-flux tension propagation theory is obtained for average translocation time as a function of rod length and self-propelling force.
Effect of active crowder size on the polymer translocation is studied in \cite{Tan} and it is found that crowders of intermediate size are most favorable for translocation in the case of forced translocation. For unbiased translocation driven by different active crowder sizes, an opposite directional preference is observed for large activities.

Here, we study polymer translocation in the presence of active particles where may be on only one or on both sides of the membrane. Initially, the middle bead of the polymer chain is fixed inside the pore and no external driving force is applied on the chain.
While in the absence of active particles, polymer translocation into either side of the membrane is equally likely, we show rectified translocation occurs by introducing active particles on one or on both sides of the membrane. 
We map out a phase diagram as a function of system parameters, such as P\'eclet numbers, area fractions and chirality, to indicate the two different regimes of rectification (namely {\em cis}-to-{\em trans} and {\em trans}-to-{\em cis} phases). We thus unveil a transition between the two regimes, where the translocation becomes excessively slow and translocation time exhibits a sharp peak at the transition. We show that the transition is produced by the interplay between effective pulling and pushing forces experienced by the polymer. 
In the case of chiral active particles we found a nonmonotonic dependence of the transition point on chirality strength.

The organization of the paper is as follows. We describe our model in Sec. \ref{model}, discuss our results in Sec. \ref{results} and conclude the paper in Sec. \ref{summary}.

\section{Model and Methods}\label{model}

Our model comprises a single flexible polymer chain consisting of $N$ dynamically passive beads with diameter $\sigma$ and mass $m_0$  linked linearly via finite extension nonlinear elastic (FENE) bonds. The latter are defined  through the stretching pair interaction energy \cite{Jin}
\begin{equation}
\label{fene}
U_{\rm FENE}(r_{ij})=-\frac{kb_0^2}{2}\ln\!\left[1-\left(\frac{r_{ij}}{b_0}\right)^2\right] 
\end{equation}
for $ r_{ij}< b_0$ and $U_{\rm FENE}=0$ otherwise, where $r_{ij}=\vert\textbf{r}_i-\textbf{r}_j\vert$ is the distance between consecutive beads labeled by $i$ and $j=1,\cdots,N$ with $|i-j|=1$, $b_0=2\sigma$ is the maximal bond stretching length,  and $k$ the bond elastic constant. The beads interact sterically among themselves via the Weeks-Chandler-Andersen (WCA) pair potential \cite{Weeks}
\begin{equation}
\label{wca}
 U_{\rm WCA}(r_{ij})= 4\varepsilon\left[\left(\frac{\sigma}{r_{ij}}\right)^{12}-\left(\frac{\sigma}{r_{ij}}\right)^{6}+\frac{1}{4}\right], 
\end{equation}
for $r_{ij}<2^{1/6}\sigma$ and $ U_{\rm WCA}=0$ otherwise, where $r_{ij}$ here is the distance between any pair of beads, and $\varepsilon$ and $2^{1/6}\sigma$  are the corresponding interaction strength and  range.

\begin{figure}[t!]
\centering
\includegraphics[width=\linewidth]{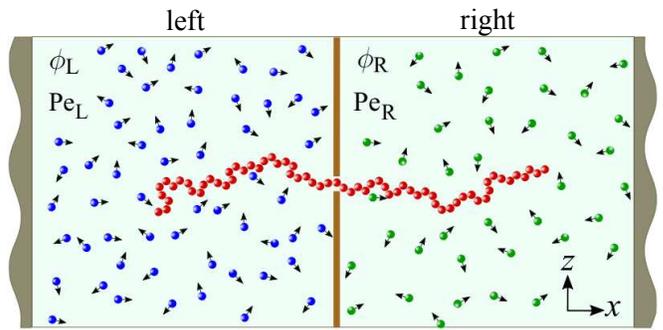}
\caption{Schematic view of a passive flexible polymer translocating through a membrane nanopore in the presence of active particles. The latter can be present either on (i) one side of the membrane at area fraction $\phi$ and P\'eclet number $\text{Pe}$ (not shown) or (ii) on both sides of the membrane with left (right) area fraction $\phi_\text{L}$  ($\phi_\text{R}$) and P\'eclet number $\text{Pe}_\text{L}$ ($\text{Pe}_\text{R}$).
}
\label{fig1}
\end{figure}

The polymer is suspended in a continuum solvent and translocates through a circular nanopore of diameter $2\sigma$ designated in a thin, rigid and impermeable membrane of thickness $\sigma$; see Fig.  \ref{fig1}. The system also contains active Brownian particles, self-propelling due to constant self-propulsive forces of magnitude $F_0$. For simplicity, we assume that the active particles have the same size and mass as the polymer beads (in the overdamped regime to be considered here, the particle mass becomes obsolete)  and that they interact among themselves and with the polymer beads via the same WCA potential as in Eq. (\ref{wca}). All particles are constrained to the $x-z$ plane within a rectangular box of dimensions $L_x$ and $L_z=L_x/2$ in $x$ and $z$ directions, respectively, with the dividing membrane positioned in the middle of the box and the nanopore at the center of the membrane (Fig. \ref{fig1}). The active particles can be present (i) only on one side of the dividing membrane, which we conventionally take as its left side and identify it later as the {\em cis}-side, or (ii) on both its left ({\em cis}) and right  ({\em trans}) sides, where the active particles can appear with unequal self-propulsive forces and area fractions (see below).

\begin{figure*}[t!]
\centering
\includegraphics[width=\textwidth]{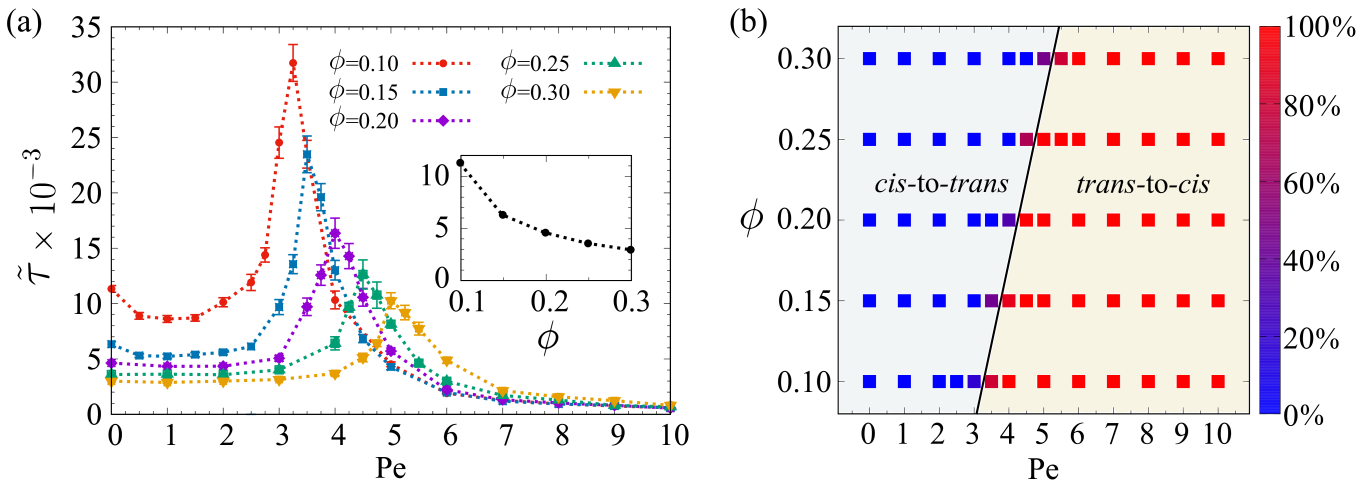}
\caption{
(a) Main set: Translocation time of a polymer chain in the presence of active particles on {\em cis}-side of the membrane as a function of particles P\'eclet number for different values of active particle area fractions. Inset: Translocation time of a polymer chain in the presence of passive particles on the {\em cis}-side as a function of particles area fraction. Symbols are simulation data and dotted curves are guides to the eye. (b) Percentage of the chain translocation into the {\em cis}-side of the membrane in the $\phi-{\rm Pe}$ phase space. Squares are simulation data and black line demonstrates the boundary between the two phases.
}
\label{fig2}
\end{figure*}

The translational dynamics of active particles are described by the Langevin equation \cite{Zwanzig} 
\begin{equation}
\label{Lan_eqtr_a}
m_0\ddot{\textbf{r}}_i=F_0\hat{\textbf{u}}_i-\frac{\partial  U}{\partial\textbf{r}_i}-\gamma_\text{tr}\dot{\textbf{r}}_i+\sqrt{2\gamma_\text{tr} k_{\rm B}T}\boldsymbol\eta_i(t),
\end{equation}
where ${\textbf{r}}_i$ is the position of the $i$th active particle and $\hat{\textbf{u}}_i$ is its orientation vector to be parametrized by an orientation angle $\varphi_i$ relative to the $x$-axis, $\hat{\textbf{u}}_i=(\cos\varphi_i,\sin\varphi_i)$, and $U$ is the total energy of particle $i$ (sum of all steric and elastic pair interactions, as described above). For the polymer beads, we use the same dynamical equation as \eqref{Lan_eqtr_a} but with $F_0=0$ as the beads  are modeled as passive Brownian particles (the polymer dynamics is thus given by the Rouse model \cite{Rubinstein}; see Section \ref{summary}). In Eq. \eqref{Lan_eqtr_a}, $\gamma_\text{tr}$ is the single-particle (bulk) translational friction coefficient of the particles and the (thermal) random forces satisfy $\langle\eta_i(t)\rangle=0$ and $\langle\eta_i^{\alpha}(t)\eta_j^{\beta}(t')\rangle=\delta_{ij}\delta_{\alpha\beta}\delta(t-t')$, where $\alpha, \beta=x, z$ indicate the Cartesian components.  In the case of active particles, the dynamics of orientation vector is modeled through the equation \cite{Zwanzig}
\begin{equation}
\label{Lan_eqrot_a}
\dot{\varphi}_i=\omega+\sqrt{\frac{2 k_{\rm B}T}{\gamma_\text{rot}}}\zeta_i(t),
\end{equation}
where $\omega$ is angular speed of active particles which is nonzero only in the case of chiral active particles. $\gamma_\text{rot}$ is the (bulk)  rotational friction coefficient and the random torque  satisfies $\langle\zeta_i(t)\rangle=0$ and $\langle\zeta_i(t)\zeta_j(t')\rangle=\delta_{ij}\delta(t-t')$. 

With further details provided in Appendix \ref{app1}, we only note here that Eqs. \eqref{Lan_eqtr_a} and \eqref{Lan_eqrot_a} are  solved numerically to obtain the  translocation time $\tau$, which is reported in rescaled units $\tilde \tau=\tau/\tau_0$ with $\tau_0=\sigma\sqrt{m_0/\varepsilon}$ taken as characteristic time scale. We start the simulations from an initial state where the middle polymer bead is manually  fixed inside the pore and the self-propulsive forces are turned off ($F_0=0$)  for all particles  as the system is equilibrated before  the self-propulsive forces are switched on and the middle bead  released. $\tau$ gives the time taken (from the moment of release) for either of the polymer end-beads to pass through the pore \cite{Huopaniemi}. We fix the number of polymer beads,  the chain flexibility and the particle radii (being equal among different particle species) and focus only on the role of particle activity, chirality and crowding that are characterized by the active particle P\'eclet number, chirality strength and area fraction. When active particles are present only on one side of the membrane (case i), the latter quantities are denoted by ${\rm Pe}$, $\Gamma$ and $\phi$, with the P\'eclet number defined standardly as 
 \begin{equation}\label{def_Pe}
{\rm Pe}=\frac{\sigma F_0}{k_{\rm B}T}. 
\end{equation}
When active particles are on both sides of the membrane (case ii; Fig. \ref{fig1}), for the sake of simplicity and reducing number of variables, we only consider nonchiral active particles. In this case, the left/right area fractions, $\{\phi_\text{L}, \phi_\text{R}\}$, and the left/right P\'eclet numbers, $\{\text{Pe}_\text{L}, \text{Pe}_\text{R}\}$, are defined accordingly; see Appendix \ref{app1}.

\begin{figure*}[t!]
\centering
\includegraphics[width=\textwidth]{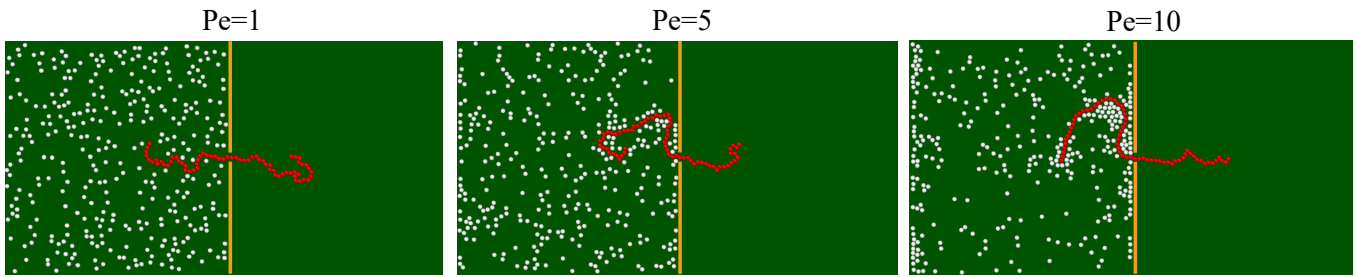}
\caption{
Typical snapshots for translocation of the polymer chain in the presence of active particles on the {\em cis}-side of the pore and with P\'eclet numbers ${\rm Pe}=1$, ${\rm Pe}=5$ and ${\rm Pe}=10$. Area fraction of active particles is taken as $\phi=0.1$.
}
\label{fig3}
\end{figure*}

\section{Results}\label{results}

\subsection{Active particles on {\em cis}-side (case i)}\label{case_i}
\subsubsection{Nonchiral active particles} 
Before proceeding with the discussion of our results for the case of nonchiral active particles (with $\Gamma=0$),  we consider the baseline case where the polymer translocation occurs in the presence of passive Brownian particles ($\text{Pe}=0$) confined to the left side of the membrane. As expected, the osmotic force that arises from the steric interactions of the polymer and the passive particles always drives out the polymer \cite{Gopinathan,Cao} with a translocation time, $\tilde \tau$,  that depends on the area fraction of the passive bath $\phi$, as shown in the inset of Fig. \ref{fig2}(a).  Taking this  process as the  reference  {\em cis-}to-{\em trans} translocation, the left side can be identified as the {\em cis-}side, even when the active self-propulsion of the bath particles are turned on (this differs from Refs. \cite{Pu,Khalilian} where the compartment containing the active particles is designated as the {\em trans-}side). 

The polymer translocation time in the presence of active particles of P\'eclet number $\text{Pe}$   on the  {\em cis-}side is shown in the main set of Fig. \ref{fig2}(a)  for different fixed area fractions of active particles, $\phi$. As seen, at sufficiently small ${\rm Pe}$, the translocation time remains nearly unchanged relative to its reference value at  $\text{Pe}=0$ albeit with a shallow minimum discernible at the smaller area fractions, $\phi\lesssim 0.15$. This means that, small activity of the particles at small area fractions facilitates translocation of the polymer chain into the {\em trans}-side and reduces the translocation time. This is because at small $\rm Pe$, activity is not large enough to make particle accumulations around the polymer chain, but instead, slight collisions of the active particles with the polymer can help it to translocate into the {\em trans}-side.
As the P\'eclet number is  increased to an intermediate value of $\text{Pe}_\ast$, the translocation time increases rapidly to a maximum value before it falls off, tending to zero as  $\text{Pe}$ is further increased. This nontrivial behavior can partially be interpreted as a consequence of prolonged detention times of active particles and their crowding near the confining walls and especially around the polymer chain, at larger P\'eclet numbers. We discuss this phenomenon with more details in the following.

In Fig. \ref{fig2}(b), percentage of the chain translocation into the {\em cis}-side of the box is plotted in the $\phi-{\rm Pe}$ phase space. Blue demonstrates that the chain always escapes from the active medium and enters the {\em trans}-side, and red indicates its entry with $100\%$ probability into the {\em cis}-side.  An interesting result from this plot is that, for small P\'eclet numbers, the polymer evades the active particles and enters the {\em trans}-side; but, by increasing $\rm Pe$, direction of the translocation is reversed and the chain translocates into the {\em cis}-side. According to this outcome, we define {\em cis}-to-{\em trans} ({\em trans}-to-{\em cis}) phase as the situation in which the chain translocates with probability $>50\%$ into the {\em trans} ({\em cis}) side. The solid line in the figure shows approximate boundary between these phases. It is beneficial to first take a look at some typical snapshots of the system.
In Fig. \ref{fig3}, snapshots are shown for three values of the active particles P\'eclet number. For small activities, the active particles have an almost uniform distribution in the box and around the chain (as seen for ${\rm Pe}=1$). At intermediate P\'eclet numbers ($\text{Pe}=5$), the active particles tend to accumulate near the walls and the polymer, and their accumulation becomes more evident at large activities ($\text{Pe}=10$).
Motivated by these results, we consider role of the effective pulling and pushing forces between the polymer and active particles. On one hand, because of persistent motion of the active particles, they spend longer times near the polymer chain and their accumulation pull the chain into the active side, so an effective pulling can be attributed to them. On the other hand, polymer is pushed by the particles due to the excluded volume effects and osmotic force. As seen in Fig. \ref{fig2}(b), at a fixed value of $\phi$ and for small P\'eclet numbers, we can almost see a same result as passive case and the polymer chain always translocates into the right side ({\em cis}-to-{\em trans} phase). By increasing $\rm Pe$ around the maximum, the active particles become stronger to capture the  chain and prevent it from escaping; here, a competition between pushing of steric effects and pulling of activity leads the polymer to make an exit from either side of the membrane. The boundary (shown as solid line in Fig. \ref{fig2}(b)) between {\em cis}-to-{\em trans} and {\em trans}-to-{\em cis} phases is where the translocation of the chain into both sides of the membrane becomes equally likely. Consequently, this competition results a maximum in the translocation time, as seen for all values of $\phi$ in Fig. \ref{fig2}(a). The maximum point can be considered as a transition point between the aforementioned phases. Finally, for large $\rm Pe$ values, activity dominates steric effects and active particles are strong enough to enforce the chain to always translocate into the active side ({\em trans}-to-{\em cis} phase). 
The observed behavior shows that, polymer translocation in active media can be rectified and its direction can be reversed under the influence of active agents. 

\begin{figure*}[t!]
\centering
\includegraphics[width=\textwidth]{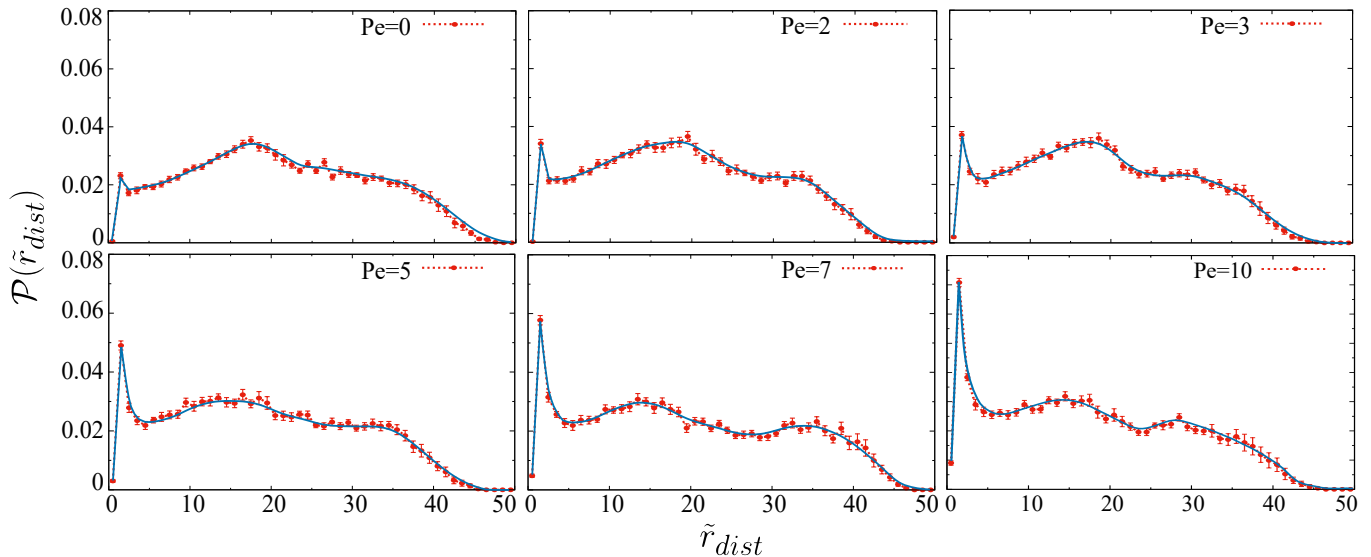}
\caption{
Distance distribution of active particles from the polymer chain (probability of finding an active particle at the specified distance $\tilde{r}_{dist}$ from the polymer) for different values of particle P\'eclet numbers. Area fraction of particles is $\phi=0.2$.
}
\label{fig4}
\end{figure*}

Higher peak at smaller values of ${\rm Pe}_\ast$ can be seen in Fig. \ref{fig2}(a) for small area fractions, while, it drops off for larger values of area fractions, with a shift toward larger ${\rm Pe}_\ast$. This result can also be observed in Fig. \ref{fig2}(b) where for larger $\phi$, the transition between {\em cis}-to-{\em trans} and {\em trans}-to-{\em cis} phases occurs at larger P\'eclet numbers. This shows that for larger area fractions where number of the active particles on the {\em cis}-side is large, although the effective pulling due to the activity grows, the pushing due to the steric effects also becomes strong; hence, only at larger P\'eclet numbers, activity can overcome steric effects and the particles can capture the polymer chain.

For a better understanding of the role played by active forces in the translocation of the polymer chain, the distribution of active particle distance from the chain, ${\cal P}(\tilde{r}_{dist})$, is plotted in Fig. \ref{fig4} for different values of particle P\'eclet number. This quantity is the probability of finding an active particle at a specified distance, $\tilde{r}_{dist}$, from the polymer chain. ${\cal P}(\tilde{r}_{dist})$ can be obtained upon appropriate averaging from the simulations and is defined as
\begin{equation}
{\cal P}(\tilde{r}_{dist})=\frac{1}{N_\text{a}}\langle n(\tilde{r}_{dist})\rangle,
\end{equation}
where $N_\text{a}$ is number of the active particles on the {\em cis}-side of the box. $n(\tilde{r}_{dist})$ is number of the active particles found at the distance $\tilde{r}_{dist}$ from the chain and $\langle\cdots\rangle$ denotes ensemble averaging.
As seen in Fig. \ref{fig4}, for passive bath (${\rm Pe}=0$), the particles move due to the Brownian motion and they are randomly distributed inside the box without any accumulation near the polymer chain. 
In this case, the most probable distance of the particles from the polymer is somewhere in the bulk of the box where can be considered about comparable to amplitude of the fluctuations of the polymer chain. Active particles tend to accumulate close to the polymer, therefore in the presence of active particles, a sharp peak in the distribution profile is generated at a distance comparable to single particle size (${\rm Pe}=2$). By increasing P\'eclet number, a large number of active particles accumulate around the chain and this sharp peak grows (as seen for ${\rm Pe}=3,5,7,10$). As a result of this dense layer around the polymer chain, active particles would be able to capture the chain and pull it into the active side of the box.

Active particles layering around the polymer chain is another point that can be inferred from Fig. \ref{fig4}. This phenomenon is a nonequilibrium effect arising from both active particles tendency for accumulation near surfaces and the repulsive steric interactions between them. Such behavior is observed as density rings in a system of inclusions in an active bath \cite{Zarif,Sebtosheikh}. As demonstrated in Fig. \ref{fig4}, there are two consecutive pronounced layers of particles in a distance relatively far from the chain which are present in all plots and can be regarded as {\em steric} layers. These layers for $\rm Pe=0$ can be related to fluctuations of the polymer chain in the passive bath. For $\rm Pe\neq 0$, these steric layers are more discernible and in addition to them, a layer of particles in the close vicinity of the polymer chain ($\tilde{r}_{dist}\simeq1$) also emerges which we can refer to it as {\em active} layer. For large P\'eclet numbers, the active layer grows and we can see an array of three layers in the distribution profile (distinguished as a blue curve with three peaks). It should be noted that the polymer chain here can be viewed as a dynamic wall or fluctuating membrane and because of its fluctuations, multiple particle layers are developed at relatively large distances from the chain. It turns out that if we replace our flexible chain with a static rigid rod, particle layers develop at smaller distances from each other and the distribution profile decays more rapidly and reaches zero at around middle of the box.

\begin{figure*}[t!]
\centering
\includegraphics[width=\textwidth]{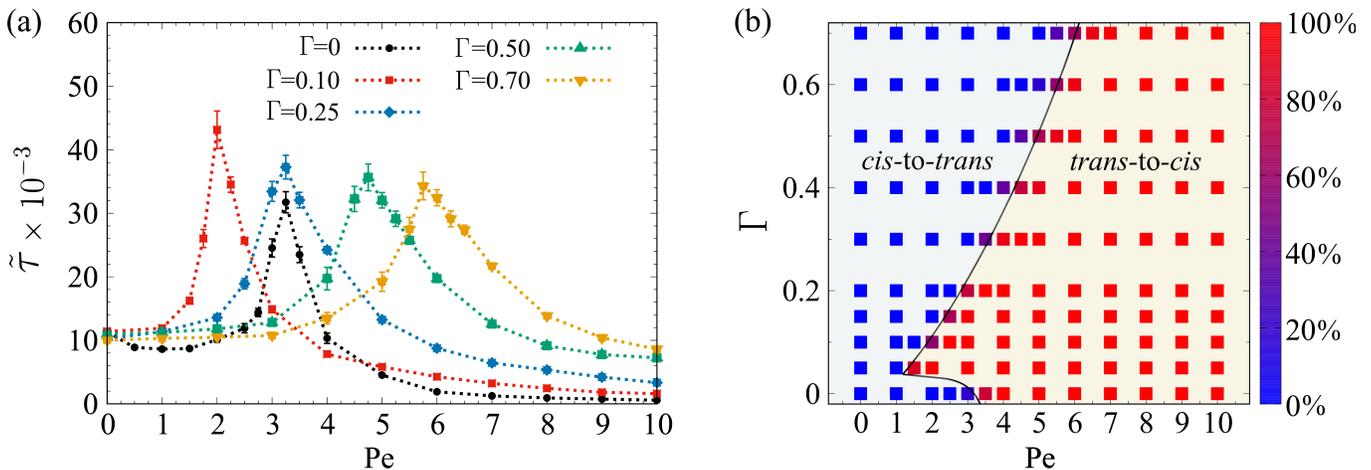}
\caption{
(a) Translocation time of a polymer chain in the presence of chiral active particles on {\em cis}-side of the membrane as a function of particle P\'eclet numbers for different fixed chirality strengths. Symbols are simulation data and dotted curves are guides to the eye. (b) Percentage of the chain translocation into the {\em cis}-side of the membrane in the $\Gamma-{\rm Pe}$ phase space. Squares are simulation data and black curves demonstrate the boundary between the two phases. Area fraction of particles in both plots is set to $\phi=0.1$.
}
\label{fig5}
\end{figure*}

\subsubsection{Chiral active particles}
Chirality can impute rotational motion to active particles and as a result, in confined systems, chirality-induced currents are generated on boundaries and surfaces \cite{Jamali,Fazli}. Here, we can incorporate chirality in our model by including a finite angular speed for active particles ($\omega$) in Eq. (\ref{Lan_eqrot_a}).
In presenting our results, we use the dimensionless angular speed or chirality strength, $\Gamma$ (see Appendix \ref{app1} for dimensionless units).

In Fig. \ref{fig5}(a), translocation time of a polymer chain in the presence of chiral active particles with area fraction $\phi=0.1$ on the left side of the membrane is plotted as a function of the particles P\'eclet number and for different fixed chirality strengths. In all plots of this figure a maximum can be seen and consequently a transition occurs. 
Here, ${\rm Pe}_\ast$ depends on the chirality strength and by increasing $\Gamma$ from zero it shows a nonmonotonic behavior. For small values of chirality ($\Gamma<0.25$), the whole curve is shifted toward left when compared with nonchiral case ($\Gamma=0$). For $\Gamma=0.25$ the transition occurs at $\text{Pe}_\ast=3.25$ which is the same value of P\'eclet number for the transition in nonchiral case. For larger chirality strengths where the active particles have larger angular speed ($\Gamma>0.25$), larger activities are required to apply a larger effective pulling force on the chain and make the transition to occur from {\em cis}-to-{\em trans} to {\em trans}-to-{\em cis} phase.
These results can be found more easily in panel (b) of Fig. \ref{fig5} where {\em cis}-to-{\em trans} and {\em trans}-to-{\em cis} phases are shown in $\Gamma$-{\rm Pe} phase space. Area fraction of active particles is fixed to $\phi=0.1$ and colors demonstrate percentage of the chain translocation into the {\em cis}-side of the box.

A remarkable point in Fig. \ref{fig5}(a) is the increase of translocation time in the presence of chiral active particles in comparison with the nonchiral case. This outcome directly results from particle chirality which weakens the persistent motion of active particles by reorienting their direction of motion and thus prevents their accumulation around the chain. Consequently, the effective pulling induced by activity becomes weaker and the translocation time increases. 

\begin{figure*}[t!]
\centering
\includegraphics[width=\textwidth]{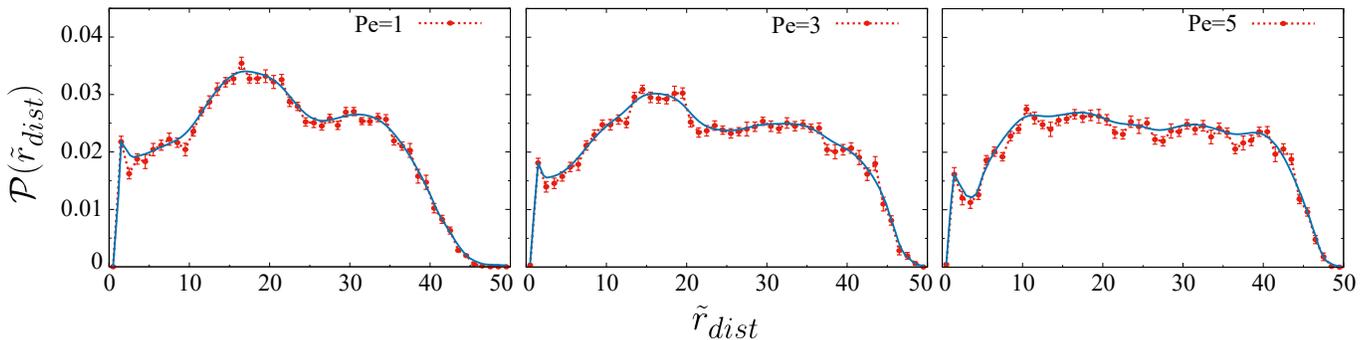}
\caption{
Distance distribution of chiral active particles from the polymer chain for three values of particle P\'eclet numbers. Chirality strength of active particles is set to $\Gamma=0.2$ and their area fraction is $\phi=0.1$.
}
\label{fig6}
\end{figure*}

The distance distribution of chiral active particles from the polymer chain is plotted in Fig. \ref{fig6} for three different values of particle P\'eclet numbers. Particles accumulation near the chain is significantly reduced in comparison with the nonchiral case (Fig. \ref{fig4}) and we can not observe dense particle layers near the chain. As stated above, this is a result of chirality and intrinsic angular velocity of the particles. Instead, particles mostly distribute far from the chain in the bulk where we can see two consecutive steric layers for small activities (${\rm Pe}=1,3$). For large P\'eclet numbers, a nearly flat distribution of particles around the chain can be found (${\rm Pe}=5$). This can be interpreted by considering the fact that, for a fixed chirality strength and by increasing P\'eclet number, chiral active particles move on the larger circles where this can affect their spatial distribution and destroys steric particle layers.

\subsection{Active particles on {\em cis} and {\em trans}-sides (case ii)}\label{case_ii}

\begin{figure*}[t!]
\centering
\includegraphics[width=\textwidth]{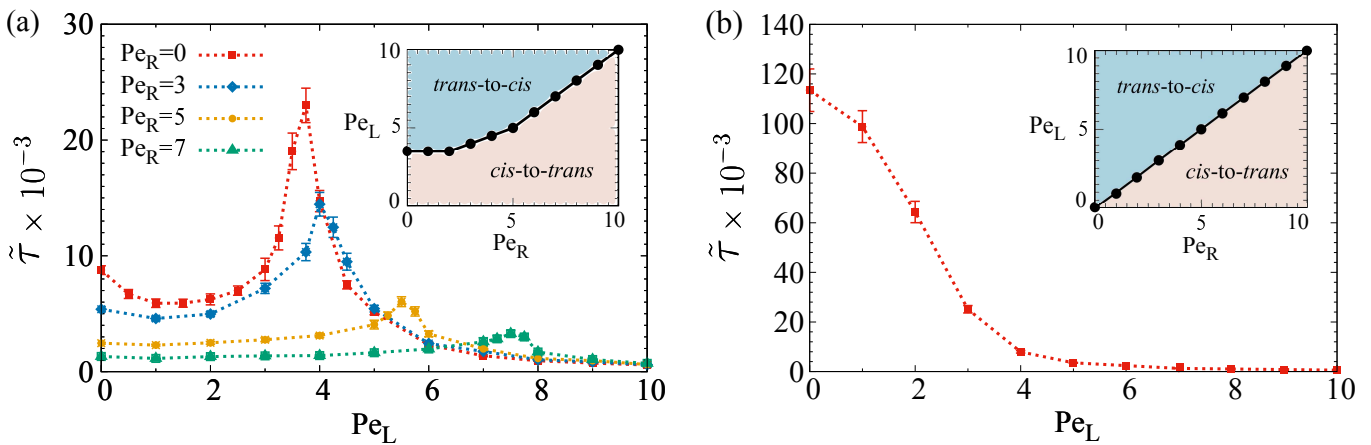}
\caption{
(a) Main set: Translocation time of a chain in the presence of active particles with area fractions $\phi_{\rm L}=0.2$ and $\phi_{\rm R}=0.1$ on the {\em cis} and {\em trans} sides of the pore, respectively. Translocation times are plotted as a function of P\'eclet number of particles on the {\em cis}-side, ${\rm Pe_L}$, and obtained for four different values of P\'eclet number of particles on the {\em trans}-side,  ${\rm Pe_R}$. Inset: {\em cis}-to-{\em trans} and {\em trans}-to-{\em cis} regimes in the (${\rm Pe_R}, {\rm Pe_L}$) phase plain.  (b) Main set: Translocation time of a chain in the presence of particles with equal area fractions $\phi_{\rm L}=\phi_{\rm R}=0.1$ on the {\em cis} and {\em trans} sides of the pore, plotted as a function of P\'eclet number of particles on the {\em cis}-side ${\rm Pe_L}$, and for ${\rm Pe_R}=0$. Inset: {\em cis}-to-{\em trans} and {\em trans}-to-{\em cis} regimes in the (${\rm Pe_R}, {\rm Pe_L}$) phase plain. Symbols are simulation data and curves are guides to the eye.
}
\label{fig7}
\end{figure*}

In this section, we study translocation of the polymer chain in the presence of nonchiral active particles on both sides of the membrane.
In the main set of Fig. \ref{fig7}(a), translocation time of a polymer chain in the presence of passive (${\rm Pe_R}=0$) and active (${\rm Pe_R}=3,5,7$) particles on the right ({\em trans}) side with area fraction $\phi_{\rm R}=0.1$ is plotted as a function of P\'eclet number of particles on the left ({\em cis}) side, ${\rm Pe_L}$. In this case, we set area fraction of the particles on the {\em cis}-side to $\phi_{\rm L}=0.2$. As in the former case (case i), the maximum in these plots denotes the transition point from {\em cis}-to-{\em trans} to {\em trans}-to-{\em cis} phase. As expected, the position of the maximum in each plot depends on the corresponding ${\rm Pe_R}$.
For ${\rm Pe_R}=0$ and at low ${\rm Pe_L}$, because of the larger area fraction and consequently dominance of the effective pushing on the {\em cis}-side, the polymer translocates into the {\em trans}-side. The shallow minimum observed at small ${\rm Pe_L}$, indicates that small particle activities on the {\em cis}-side facilitate translocation of the polymer into the {\em trans}-side. 
For larger ${\rm Pe_L}$, at the transition point (around the maximum), as a result of the competition between the effective pulling and pushing on both sides, all plots in Fig. \ref{fig7}(a) show an increase of translocation time. Finally, for large ${\rm Pe_L}$, pulling due to activity on the {\em cis}-side overcomes both effective pushing on the {\em cis}-side and pulling on the {\em trans}-side, and as a result, the chain translocates into the {\em cis}-side of the membrane and the translocation time decays. 
An interesting point in this figure is that for large ${\rm Pe_L}$, only small differences can be seen between the four plots. This means that, for large enough activities, the effective pulling on the {\em cis}-side is the dominant factor on the translocation speed and other factors have low impacts.

Inset of Fig. \ref{fig7}(a) represents {\em cis}-to-{\em trans} and {\em trans}-to-{\em cis} regimes in the ${\rm Pe_L}$-${\rm Pe_R}$ phase space for {\em cis} and {\em trans} side area fractions $\phi_{\rm L}=0.2$ and $\phi_{\rm R}=0.1$. This figure illustrates that, by increasing ${\rm Pe_L}$ and for a fixed value of ${\rm Pe_R}$, we always have a transition from {\em cis}-to-{\em trans} to {\em trans}-to-{\em cis} phase. On the other hand, if we fix ${\rm Pe_L}$ and increase ${\rm Pe_R}$, at small values of ${\rm Pe_L}$, we can not observe any transition. This results from the fact that for small ${\rm Pe_L}$ effective pulling due to the activity on the {\em cis}-side is weak and not sufficient to overcome effective pushing, thus, due to the larger area fraction of the particles on the {\em cis}-side, effective pushing on the {\em cis}-side dominate and the chain always translocates into the {\em trans}-side. For larger values of ${\rm Pe_L}$, as a result of the stronger pulling on the {\em cis}-side, a transition from {\em trans}-to-{\em cis} to {\em cis}-to-{\em trans} phase is observed by increasing ${\rm Pe_R}$.

In Fig. \ref{fig7}(b), translocation time in the presence of passive particles on the {\em trans}-side and active particles on the {\em cis}-side with equal area fractions ($\phi_{\rm L}=\phi_{\rm R}=0.1$) is plotted as a function of ${\rm Pe_L}$. Translocation time shows a decreasing trend for all values of ${\rm Pe_L}$ with a very fast decay at small ${\rm Pe_L}$. In contrast with the former case where $\phi_{\rm L}\neq\phi_{\rm R}$ (Fig. \ref{fig7}(a)), no maximum is observed here for ${\rm Pe_L}\neq 0$. In this case, equality of area fractions on the two sides results equal effective pushings which can offset each other, therefore, at very small ${\rm Pe_L}$ where active forces are not strong enough to break the symmetry of the two sides, the translocation time takes very large values. In fact, due to the equal area fractions on the two sides of the membrane in this case and passive particles on the {\em trans}-side, the maximum point is located at ${\rm Pe_L}=0$. By activating particles on the {\em cis}-side and increasing their P\'eclet number, the pulling force on the {\em cis}-side would be the main effective mechanism on the translocation and it helps the polymer to rapidly translocate through the pore. 
It turns out that for ${\rm Pe_R}\neq0$, the transition point (maximum) shifts toward ${\rm Pe_L}\neq0$ (as expected), and behavior of the translocation time is qualitatively similar to what we observed in Fig. \ref{fig7}(a). This is an immediate result of breaking the symmetry  of the effective forces acting on the two sides by activating the particles on the {\em trans}-side. In the inset of Fig. \ref{fig7}(b), {\em cis}-to-{\em trans} and {\em trans}-to-{\em cis} phases corresponding to $\phi_{\rm L}=\phi_{\rm R}=0.1$ are plotted in ${\rm Pe_L}$-${\rm Pe_R}$ phase plain. As demonstrated, because of equal area fraction of active particles on the two sides, the transition points are located on ${\rm Pe_L}={\rm Pe_R}$ line.

\section{Concluding remarks}\label{summary}

In this work, we studied translocation of a linear flexible polymer chain through a nanopore in the presence of active particles are assumed to be on one or both sides of the membrane. Due to the persistent motion of active particles and their accumulation around the polymer chain, an effective pulling force can be applied on the chain from active particles. An effective pushing force can also be induced as a result of excluded volume repulsions between the polymer and particles. Translocation of the polymer chain into each side of the membrane is assisted by activity of the particles on that side and hindered by steric effects of them, and depending on interplay between these factors on both sides and dominance of either of them, rectification of translocation is expected. As a result of imbalance in these effective pulling and pushing, we found a transition between two rectified states, namely {\em cis}-to-{\em trans} and {\em trans}-to-{\em cis} phases, where by varying particles activity, chirality strength or their area fraction, the polymer changes its destination from one side to the other side of the pore. The translocation process becomes excessively slow at the transition and a pronounced maximum in translocation time is observed.

For the case of active particles only on one side of the membrane, we found that for larger area fraction of active particles, the transition occurs at larger activities. This is a result of stronger pushing force applied on the chain and consequently a larger value of activity is needed to overcome this pushing and prevent the chain from escaping.
On the other hand, translocation time decreases for larger area fractions, which is a result of accelerating the translocation by applying stronger pulling and phshing forces on the chain.

When active particles are on both sides of the membrane, we have more complex situations. Depending on the area fraction and activity of the particles on the two sides, the transition may or may not occur. A balance between effective pulling and pushing on the {\em cis} and {\em trans} sides results in increase of translocation time and increasing asymmetry between the two sides accelerates the translocation process.

We also observed active particles layering around the polymer chain. Because of fluctuating motion of the chain, particle layers develop at larger distances from the chain, when compared with a rigid rod. Interestingly, a thin layer of active particles with thickness about one particle size develops very close to the chain, and it gets more denser by increasing particle activity.

For the case of active particles on the {\em cis}-side of the membrane we studied effect of the chirality on the polymer translocation. In the parameter space $\Gamma$-{\rm Pe}, by increasing the chirality strength for a finite interval of P\'eclet numbers, we twice have the transition.
We also found that, it takes longer times for the polymer chain to translocate in the presence of chiral active particles in comparison with the nonchiral case and this results from chirality which reduces active particles accumultion around the chain and weakens the effective pulling and pushing forces applied on the chain.

Chiral active particles mostly distribute in the bulk and they exhibit small accumultion near the polymer chain. For a fixed chirality strength, at small particle activities, steric particle layers can be found, but for large P\'eclet numbers, particles have almost uniform distribution in the bulk.

In the problem at hand, we consider the polymer chain as Rouse chain and the active particles at the level of minimal model of active Brownian particles \cite{Romanczuk}. Hence, we neglect several important factors that could be included for a more comprehensive analysis in the future. These include the effect of hydrodynamic interactions that may play a crucial role in the polymer translocation in active media. An interesting problem would be translocation dynamics of a Zimm chain \cite{Rubinstein}  in a suspension of dipolar microswimmers \cite{Lauga}. Another factors to be investigated could be roles of flexibility of the chain and quality of the solvent. 
Interactions of the active particles with the pore or membrane can also be considered, paving the way for more intriguing scenarios to be explored in the future.

\section{Acknowledgements}
We thank Turin Cloud Services and School of Nano Science of the Institute for Research in Fundamental Sciences (IPM) for computational resources.

\appendix
\section{Choice of parameter values}\label{app1}
We choose the reference scales of length, energy and mass as $\sigma$, $\varepsilon=k_{\rm B}T$ and $m_0$. 
For case i (see \ref{case_i}), number of active particles is denoted by $N_\text{a}$, and neglecting the relatively small area occupied by the membrane, their area fraction is given by $\phi={\pi N_\text{a}\sigma^2}/(2L_xL_z)$. Using definition of P\'eclet number in the main text \ref{def_Pe}, active particles P\'eclet number is denoted by $\text{Pe}$, with associated self-propulsive force magnitude $F_0$. In this case, chirality strength is rescaled as $\Gamma=\omega\tau_0$. For case ii (see \ref{case_ii}), denoting by $N_\text{a,L}$ and $N_\text{a,R}$ number of active particles on {\em cis} and {\em trans} sides of the membrane and giving their self-propulsive forces by $F_{0,\text{L}}$ and $F_{0,\text{R}}$, the corresponding area fractions and P\'eclet numbers, defined with similar relations to the former case, are respectively given by $\{\phi_{\text{L}},\phi_{\text{R}}\}$ and $\{\text{Pe}_\text{L},\text{Pe}_\text{R}\}$.

The simulation box is periodic in the $z$ direction and bounded in the direction of $x$ by planar rigid walls at $x=0$ and $L_x$. The interactions between walls and membrane with active particles and polymer beads are modeled by the same WCA potential as in Eq. (\ref{wca}). Thus, while  the membrane remains effectively impermeable to the particles (with a choice of sufficiently large $\varepsilon$), the polymer (and no active particles) can freely translocate through the pore.

We solve the equations of motion using the ESPResSo package \cite{Weik} by employing the velocity Verlet algorithm with timesteps $0.01\tau_0$. In the numerical implementation, the active particles and  polymer beads are modeled as three-dimensional objects being constrained to 2D.   
 
To determine the time of polymer translocation through the pore, $\tau$, or $\tilde \tau=\tau/\tau_0$ in rescaled units, we start the simulations from an initial state where the middle bead of the polymer is artificially fixed inside the pore and equilibrate the system for $10^6$ timesteps by treating all particles as passive ones ($F_0=0$). We then turn on the self-propulsive force of the active particles and release  the middle bead. The translocation time is measured as the time taken (from the moment of release) for either of the polymer end-beads to pass through the pore. Each point in our plots is obtained by averaging over 100 independent runs and errorbars are obtained using standard error of data.

In performing our simulations, we fix $N=65$ (for other values of $N$, we do not expect our results to qualitatively be changed, and only quantitative variations e.g., in translocation times are expected), $L_x/2=L_z=50\sigma$, $k=70k_{\rm B}T/\sigma^2$, $\gamma_\text{tr}=10m_0/\tau_0$, $\gamma_\text{rot}=10m_0\sigma^2/\tau_0$ and vary other system parameters with a range representative values ($i.e., {\rm Pe}=0.5-10$, $\phi=0.1-0.3$, $\Gamma=0.05-0.7$). Corresponding to area fractions considered here, we change number of particles in cases i and ii in the range $N_{\rm a},N_{\rm a,L},N_{\rm a,R}=333-1000$. Our dimensionless parameters can be mapped to some realistic cases. For example, for a coarse-grained chain of Nucleic acid with its beads of the size $\sigma=1\,{\rm nm}$ and mass $m_0=10^{-22}\,{\rm kg}$, we will have a characteristic time scale of simulation about $\tau_0\simeq 0.16\,{\rm ns}$ and translational and rotational frictions about $\gamma_\text{tr}\simeq29m_0/\tau_0$ and $\gamma_\text{rot}\simeq38m_0\sigma^2/\tau_0$, respectively.
The range of P\'eclet numbers used here to discuss the representative behavior of the system can be compared with numerous examples of artificial nanoswimmers including Janus nanoparticles (${\rm Pe}\simeq 1-3$) \cite{Lee}, magnetic multi-link nanoswimmers and nanofishes  (${\rm Pe}\simeq 10^2-10^3$) \cite{Jang,Li} and platinum-loaded stomatocyte nanomotors (${\rm Pe}\simeq 1-5$) \cite{Wilson}.
As an example, for nanoswimmers in the form of chains ($l=2.78\,{\rm \mu m}$ in length and $w=0.9\,{\rm nm}$ in width) and fabricated from magnetic self-assembly of nanoparticles \cite{Cheang}, a diffusion coefficient about $D_T\simeq0.28\,{\rm \mu m^2/s}$ is observed. For this example, using relation ${\rm Pe}=vl/D_T$ and in the range of P\'eclet numbers used here, self-propulsion speeds in the range $v\simeq 0.1-1\,{\rm \mu m/s}$ can be obtained.

Our investigated range of chiralities can also be mapped to artificial active particles such as curved self-propelled rods 
($|\Gamma|\simeq 0.1-0.5$) \cite{Takagi,Takagi_2}  and self-assembled rotors ($|\Gamma|\simeq 0.1-1$) \cite{Wykes}.

\bibliography{references}

\end{document}